\documentclass[conference]{IEEEtran}
\usepackage[left=0.62in,right=0.62in,top=0.75in,bottom=1in]{geometry}
\usepackage{graphicx}
\usepackage{cite}
\usepackage{amsmath}
\usepackage{amsfonts}
\usepackage{xcolor}
\usepackage{siunitx}
\usepackage{subcaption}
\usepackage{mathtools, cuted}
\usepackage{verbatim}
\usepackage{float}

\usepackage{bbding}
\makeatletter

\def\ps@IEEEtitlepagestyle{%
  \def\@oddfoot{\mycopyrightnotice}%
  \def\@evenfoot{}%
}
\def\mycopyrightnotice{%
  {\footnotesize\hfill 978-1-7281-8510-1/20/\$31.00 \copyright2020 IEEE\hfill}%
  \gdef\mycopyrightnotice{}
}

\begin{document}

\title{Design of an Enhanced Reconfigurable Chaotic Oscillator using  G\textsuperscript4FET-NDR  Based Discrete Map}

\author{
  \IEEEauthorblockN{
    Md Sakib Hasan\IEEEauthorrefmark{1}\Envelope, Aysha S. Shanta\IEEEauthorrefmark{2}, Partha Sarathi Paul\IEEEauthorrefmark{1}, Maisha Sadia\IEEEauthorrefmark{1}, Md Badruddoja Majumder\IEEEauthorrefmark{2} and \\ Garrett S. Rose\IEEEauthorrefmark{2}
    \\
  }
  \IEEEauthorblockA{\IEEEauthorrefmark{1}Department of Electrical and Computer Engineering University of Mississippi, University, MS 38677 USA
\\
}
  \IEEEauthorblockA{\IEEEauthorrefmark{2}Department of Electrical Engineering \& Computer Science, University of Tennessee, Knoxville, TN 37996 USA
\\
}
  \IEEEauthorblockA{\Envelope Corresponding Author, mhasan5@olemiss.edu
}
  
}


%


\maketitle


\begin{abstract}
 In this paper, a novel chaotic map is introduced using a voltage controlled  negative differential resistance (NDR) circuit composed of an $n$-channel and a $p$-channel silicon-on-insulator (SOI) four-gate transistor (G\textsuperscript4FET). The multiple gates of the G\textsuperscript4FET are leveraged to create a discrete chaotic map with three bifurcation parameters. The three tunable parameters are the gain of a transimpedance amplifier (TIA), top-gate voltage of $n$-channel G\textsuperscript4FET, and top-gate voltage of $p$-channel G\textsuperscript4FET. Two methods are proposed for building chaotic oscillators using this discrete map. The effect of altering bifurcation parameters on chaotic operation is illustrated using bifurcation diagrams and Lyapunov exponent. A design methodology for building flexible and reconfigurable logic gate is outlined and the consequent enhancement in functionality space caused by the existence of three independent bifurcation parameters is demonstrated and compared with previous work. 
\end{abstract}


\section{Introduction}

 As technology scaling has slowed down over the past decade, it has become increasingly difficult to integrate more transistors in a given area. Researchers are looking into new approaches to attain better performance without increasing the number of transistors. A possible solution is increasing the number of computations that a device can perform using chaotic operation \cite{kia2017integrated}. The rich inherent dynamics of the chaotic circuits have been studied by researchers to build flexible and reconfigurable digital gates \cite{kia2015nonlinear}. On the other hand, G\textsuperscript4FET, a silicon-on-insulator (SOI) device with four independent gates, is also a promising candidate for accomplishing this goal since it has been used in various analog and digital applications with reduced transistor counts\cite{akarvardar2006four}, \cite{akarvardar2005novel}. In this work, we propose a scheme to combine the advantages of a multi-gate transistor with the large functionality space of chaotic circuits.  


Traditional discrete chaotic maps such as logistic, tent, sine map, etc. are idealized mathematical functions that require a significant number of transistors for hardware implementation. In order to reduce the hardware cost, researchers have made an effort to come up with simpler transistor-level discrete maps with good chaotic properties \cite{kia2017integrated}, \cite{dudek2003compact}. Recently, researchers have been exploring how to leverage the large functionality space provided by these discrete map chaotic circuits for security applications\cite{hua2015dynamic}, \cite{hasan2020chaos}. Most chaotic maps have only one bifurcation parameter where the space is extended by increasing the number of iterations. However, due to the system's susceptibility to noise, the reliability decreases with an increase in the number of iterations. Hence, it is highly desirable to develop a chaotic map with multiple independent control parameters which can reliably expand the functionality space with fewer number of iterations and relatively simple hardware implementation.

In this work, a voltage controlled negative differential resistance (VC-NDR) circuit consisting of two G\textsuperscript4FETs and a transimpedance amplifier is used to build a novel chaotic map with three bifurcation parameters. 
Then two methods are proposed for building chaotic oscillators using this discrete map. The excellent property of the chaotic map is illustrated with the help of bifurcation diagram and Lyapunov exponent which are plotted for different values of control parameters. It is also shown that the availability of three independent bifurcation parameters can be used to exponentially increase the functionality space.

The paper is organized as follows: Section \ref{sec_map} introduces the proposed chaotic map built with G\textsuperscript4FET  NDR circuit. Section \ref{sec_osc} elaborates on the design of chaotic oscillator and shows the resulting bifurcation diagrams and  Lyapunov exponent. A methodology for generating logic functions from chaotic oscillator is shown in section \ref{sec_logic} and the resulting enhancement in design space are explored in section \ref{sec_func_space}.
Finally, section \ref{sec_con} concludes the paper.


\section{Proposed Chaotic Map using G\textsuperscript4NDR}
\label{sec_map}
\subsection{G\textsuperscript4FET}
\begin{figure}
   \centering
   \includegraphics[scale=0.40]{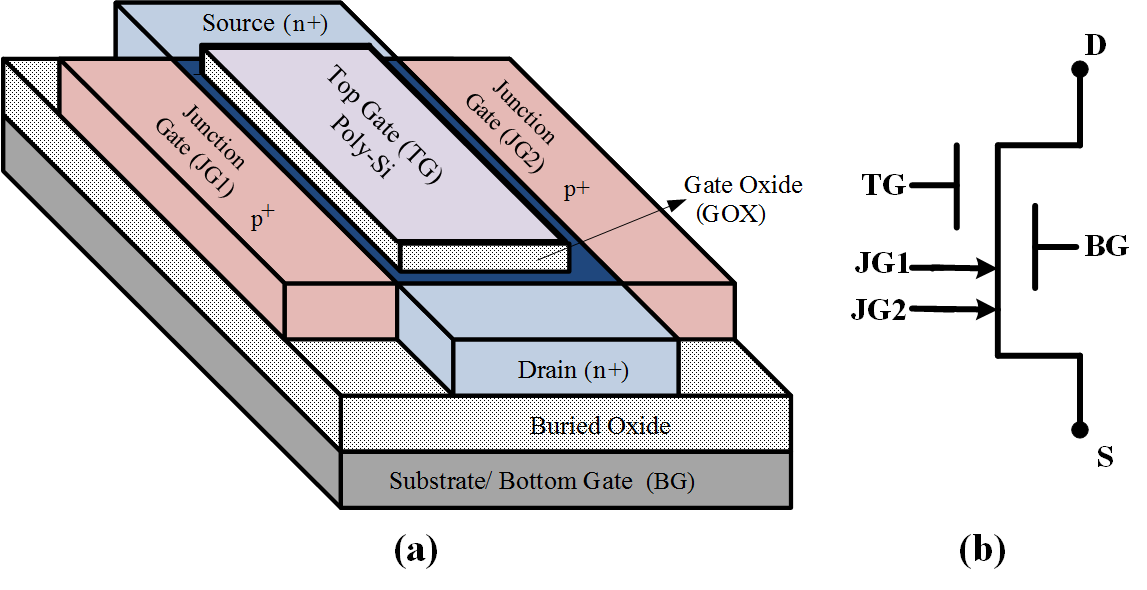}
   \setlength\belowcaptionskip{-15.0pt}
    \caption{\small{a) 3-D device structure of an $n$-channel G$^4$FET, b) circuit symbol.}}
    \label{fig:G4FET_struc}
\end{figure}

 G\textsuperscript4FET is a multigate SOI device \cite{blalock2002multiple} that is a potential candidate for designing innovative circuits with more functions and fewer transistors \cite{akarvardar2006four}, \cite{akarvardar2005novel}. It can be fabricated in partially/fully-depleted SOI process and the channel conductance can be modulated using four independent gates\cite{dufrene2004investigation}. 
 A conventional \textit{p}-channel SOI MOSFET with two body contacts on the opposite sides of the channel can be converted into an \textit{n}-channel G\textsuperscript4FET. The structure and circuit symbol of an \textit{n}-channel G\textsuperscript4FET is shown in Fig. \ref{fig:G4FET_struc}. The \textit{p}+ doped source and drain of the traditional MOSFET now operate as junction gates. The function of the two lateral gates is similar to JFET gates and can be used to control the channel width. The operation of the top gate is similar to the conventional MOS gate and the substrate is biased using the buried oxide layer which works as the bottom gate. The conductance of G\textsuperscript4FET is controlled by two MOSFET and two JFET gates that are biased independently. In this work, the bottom gates are grounded and not shown in the subsequent figures for simplicity.

\subsection{G\textsuperscript4FET NDR}
\begin{figure}
   \centering
   \includegraphics[scale=0.40]{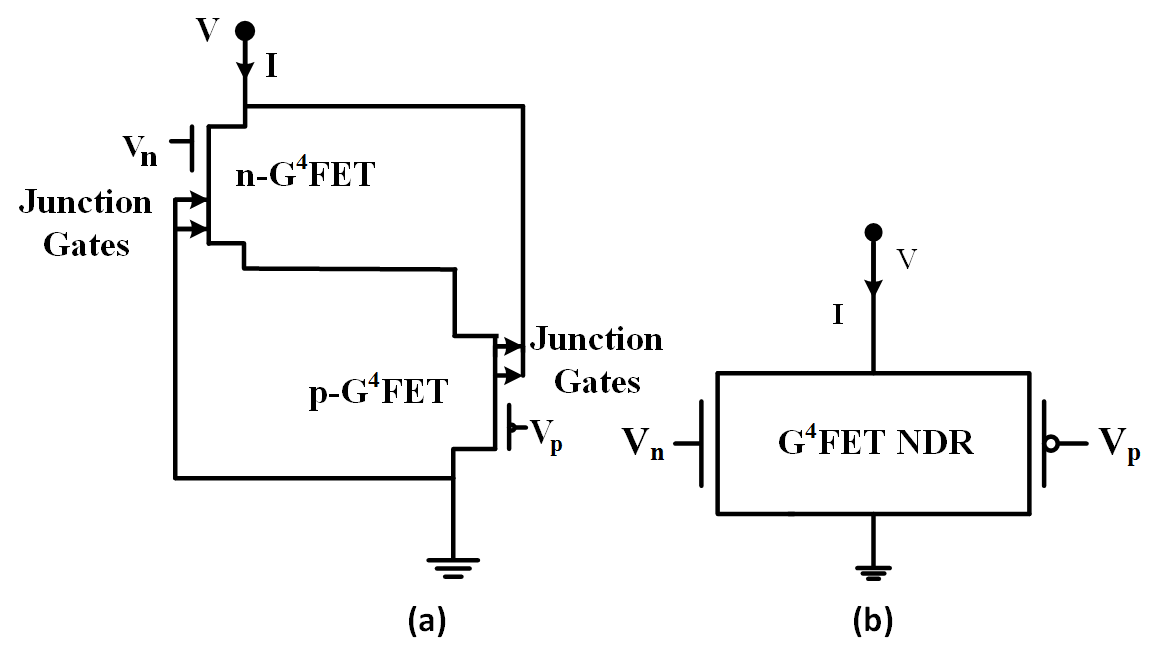}
   \setlength\belowcaptionskip{-15.0pt}
    \caption{\small{Tunable NDR circuit using $n$- and $p$-channel G$^4$FET; (a) circuit schematic, (b) simplified symbol.} }
    \label{fig:G4NDR}
\end{figure}

 A tunable VC-NDR circuit using complementary G\textsuperscript4FETs was first proposed and experimentally verified in \cite{akarvardar2006four}. The G\textsuperscript4FET NDR is developed by substituting complementary JFETs with G\textsuperscript4FETs in a conventional lambda diode\cite{takagi1975complementary}. The schematic and simplified symbol are shown in Fig. \ref{fig:G4NDR}  where both junction-gates of G\textsuperscript4FET are tied together and treated as a single gate. The G\textsuperscript4FET NDR is a four-terminal device where the additional two-terminals are the top-gates of the $n$- and $p$-channel G\textsuperscript4FETs and denoted by voltages $V_N$ and $V_P$, respectively. The dimensions of $p-$ and $n-$channel G\textsuperscript4FET in this work are $0.35$ $\mu$m / $1.2$ $\mu$m and  $0.35$ $\mu$m / $3.4$ $\mu$m respectively. All results are generated using the MOS-JFET macromodel\cite{hasan2018macromodel}.

\subsection{G$^4$NDR Based Chaotic Map (GNM)}
\begin{figure}
\centering
\includegraphics[scale=0.60]{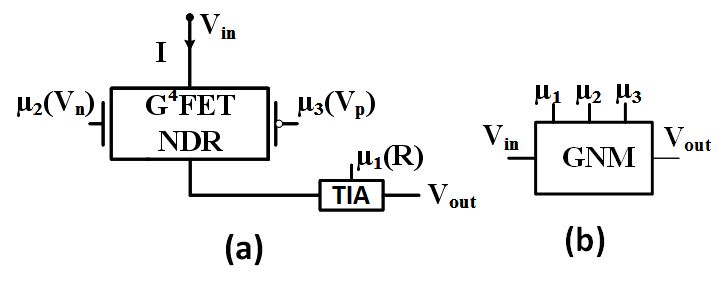}
\caption{\small{G\textsuperscript4FET NDR based map (GNM) circuit with three bifurcation parameters; (a) circuit schematic, (b) simplified symbol.} }
\label{map}
\end{figure}
\begin{figure}
   \centering
   \includegraphics[scale=0.18]{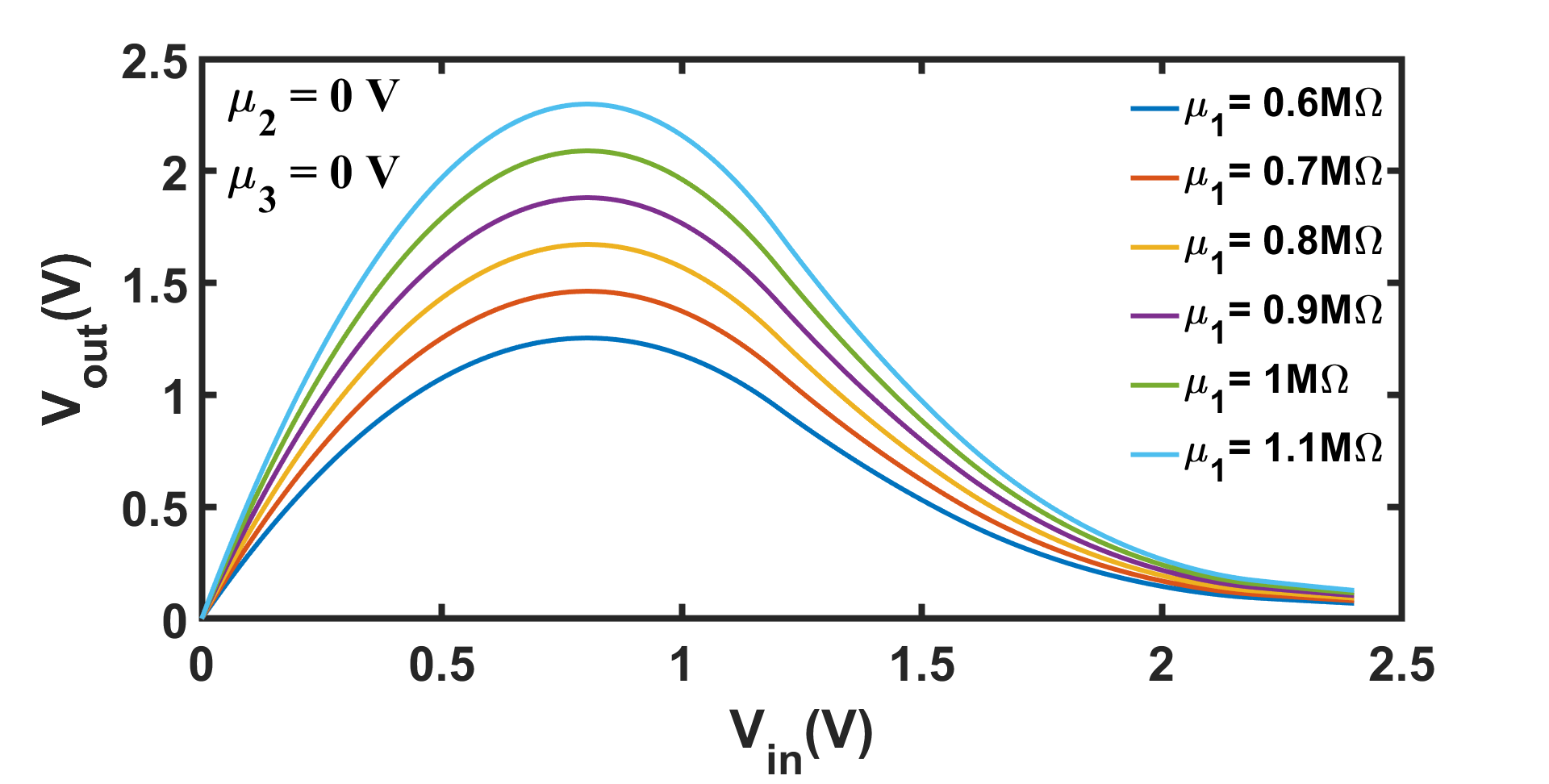}
   \setlength\belowcaptionskip{-15.0pt}
    \caption{\small{Transfer curve of the nonlinear map circuit.}}
    \label{fig:tc}
\end{figure}

If a non-linear function, $f$ transforms any point $x_n$ from a closed interval $l=[a,b]$ into some point $x_{n+1}$ in the same interval i.e. $f:l\rightarrow l$, then 
\begin{equation}
    x_{n+1}=f(x_n)
    \label{eq_disc_map}
\end{equation}

is called a discrete map of the interval $l$ for $n=1,2,...,n$. Fig. \ref{map} shows a schematic and a symbol of the chaotic map based on VC-NDR using complementary G\textsuperscript4FETs. When the resulting current is converted to voltage using a transimpedance amplifier (TIA) with a variable gain, R ($\mu_1$), we get a map circuit with transfer characteristic similar to a logistic map. However, common map functions (logistic, tent, sine etc.) are idealized mathematical functions which are expensive to implement in hardware and they provide only one bifurcation parameter. In contrast, the proposed chaotic map has three independently controllable bifurcation parameters. The transfer curve for the proposed chaotic map is shown in Fig. \ref{fig:tc} where $\mu_2$ and $\mu_3$ are set at $0$ $V$ and $\mu_1$ is varied from $0.6$ $M\Omega$ to $1.1$ $M\Omega$. The transfer curve displays differentiable unimodal characteristic which is suitable for chaotic operation.

\begin{figure}
\centering
\includegraphics[scale=0.65]{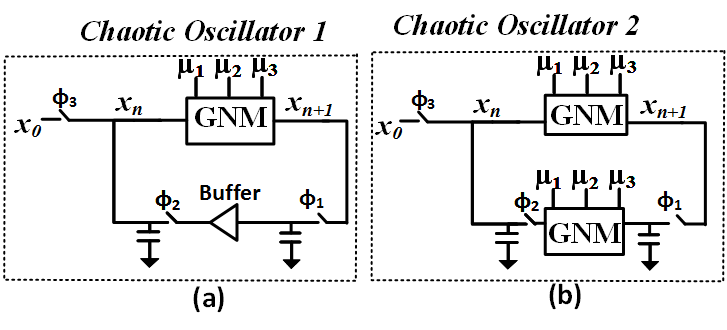}
\caption{\small{Chaotic oscillator  using G\textsuperscript4FET NDR based chaotic map (GNM); a) feedback using buffer, b) feedback using another GNM instead of the buffer.}}
\label{osci}
\end{figure}

\section{Chaotic oscillator}
\label{sec_osc}
The chaotic oscillator in Fig. \ref{osci}a is made using the G\textsuperscript4FET NDR-based chaotic map (GNM), two sample-and-hold circuits and a buffer. In Fig. \ref{osci}b, we propose another design where the buffer in the feedback loop is replaced with another map. The initial input is applied to the oscillator using clock $\phi_3$. After the initial input has been applied, $\phi_3$ stops and the first output is fed back to the input using two complementary switches $\phi_1$ and $\phi_2$. The next output states are generated by alternating the on and off condition of the switches $\phi_1$ and $\phi_2$. When the system is in the chaotic region, a new analog voltage is generated at each iteration. 

Lyapunov exponent, $\lambda$ is used to quantify the sensitive dependence of the chaotic oscillator on initial conditions. Positive values of $\lambda$ indicate chaotic operation whereas negative values imply periodic operation. For discrete-time chaotic maps, it is defined as,
 \begin{equation}
    \lambda=\lim_{n \to \infty} \frac{1}{n}\sum_{i=0}^{n-1}ln|f'(x_i)| .
    \label{eq:overhead1}
\end{equation}

\begin{figure}
    \centering
     \begin{subfigure}[b]{0.4\textwidth}
       \includegraphics[width=\textwidth]{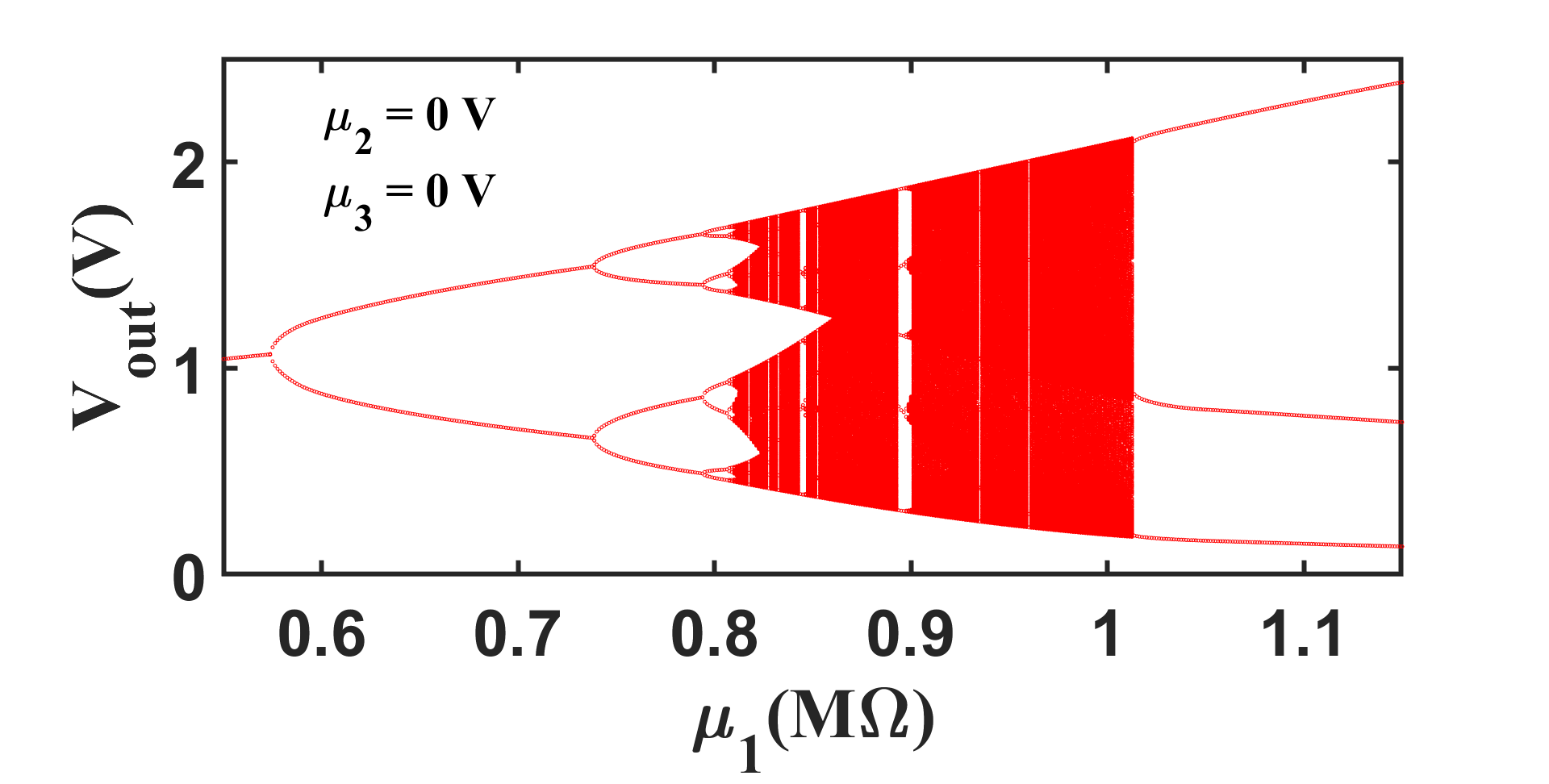}
        \caption{}
        \label{fig:bif_1}
    \end{subfigure}
    \begin{subfigure}[b]{0.4\textwidth}
      \includegraphics[width=\textwidth]{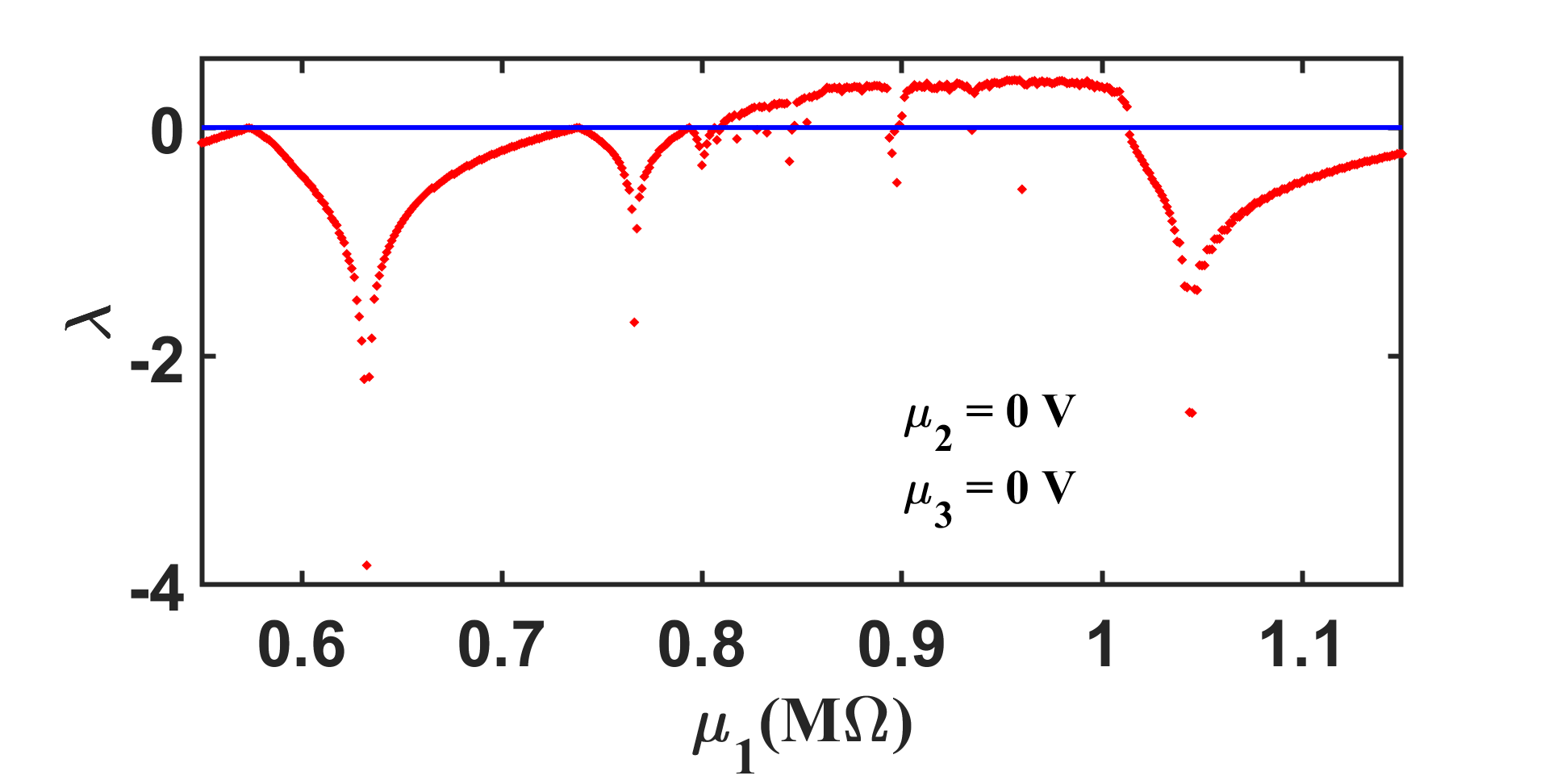}
        \caption{}
        \label{fig:LE_1}
    \end{subfigure}
    
    \caption{\small{(a) Bifurcation diagram and (b) Lyapunov exponent of the chaotic oscillator for varying $\mu_1$; $\mu_2$ = $0$ $V$,  $\mu_3$ = $0$ $V$.} }
 \label{bif_mu1}
\end{figure}

\begin{figure}
    \centering
     \begin{subfigure}[b]{0.4\textwidth}
       \includegraphics[width=\textwidth]{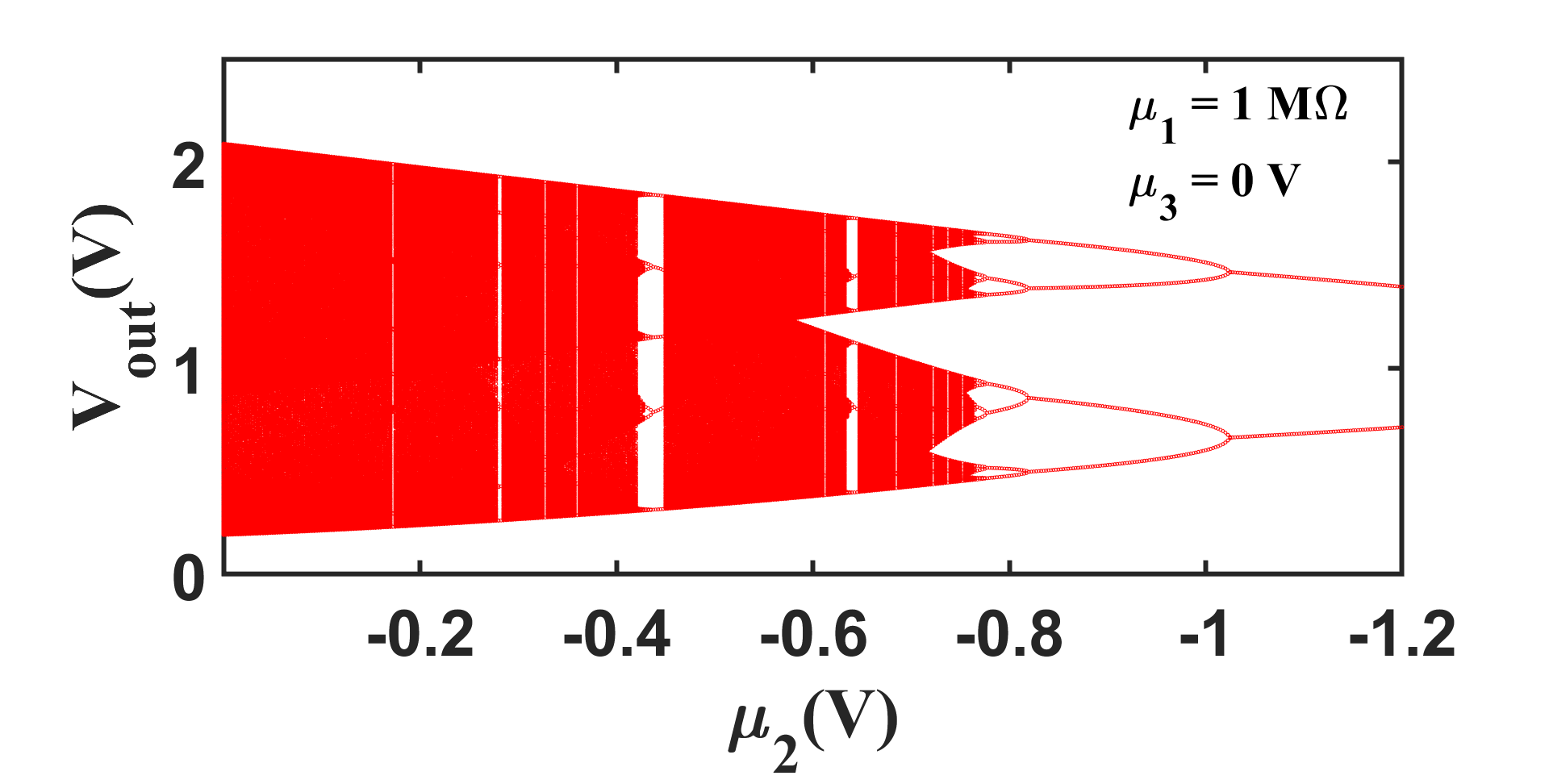}
        \caption{}
        \label{fig:bif_2}
    \end{subfigure}
    \begin{subfigure}[b]{0.4\textwidth}
      \includegraphics[width=\textwidth]{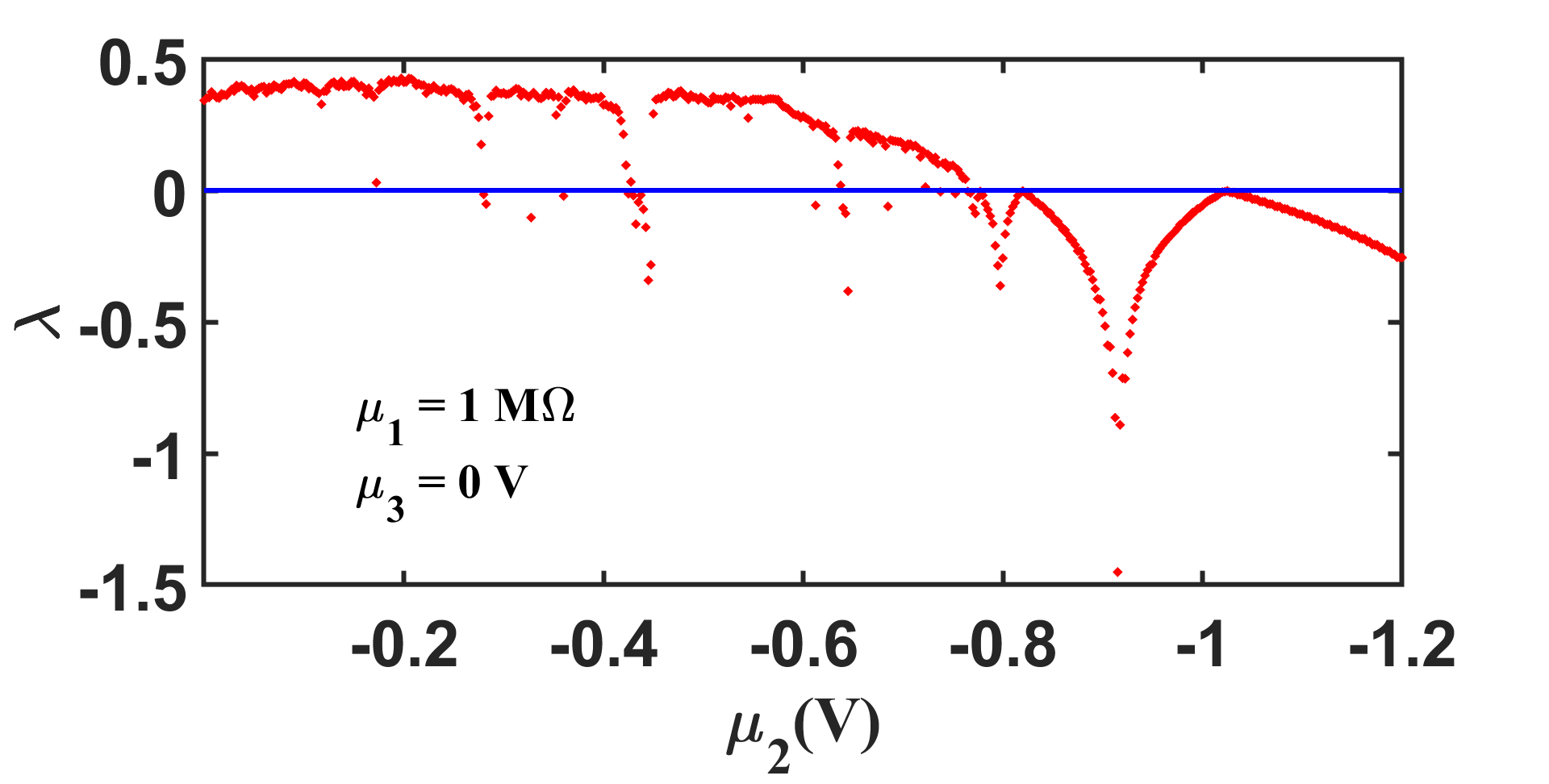}
        \caption{}
        \label{fig:LE_2}
    \end{subfigure}
    
    \caption{\small{(a) Bifurcation diagram and (b) Lyapunov exponent of the chaotic oscillator for varying $\mu_2$; $\mu_1$ = $1$ $M\Omega$,  $\mu_3$ = $0$ $V$.} }
 \label{bif_mu2}
\end{figure}

\begin{figure}
    \centering
     \begin{subfigure}[b]{0.4\textwidth}
       \includegraphics[width=\textwidth]{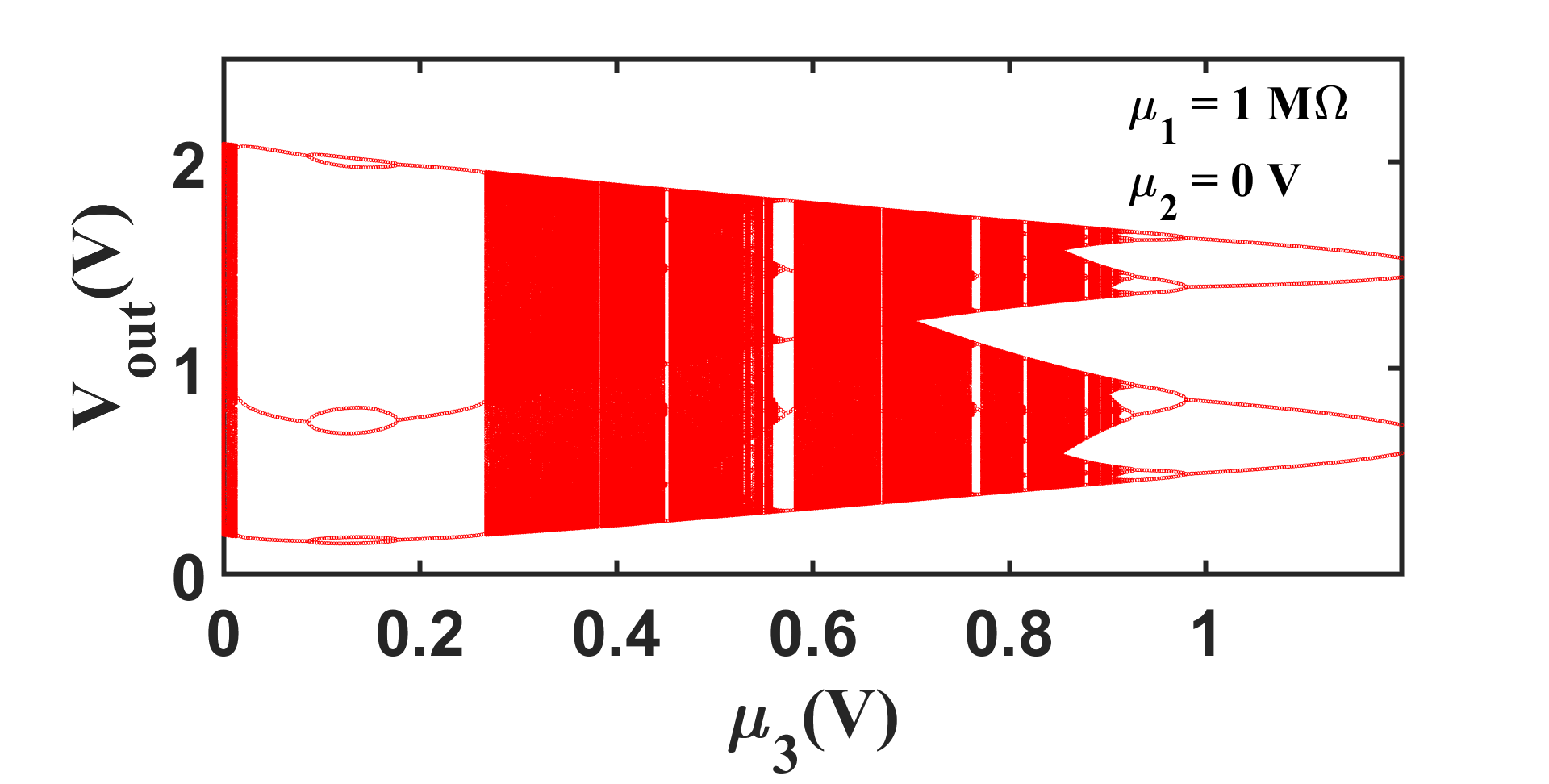}
        \caption{}
        \label{fig:bif_3}
    \end{subfigure}
    \begin{subfigure}[b]{0.4\textwidth}
      \includegraphics[width=\textwidth]{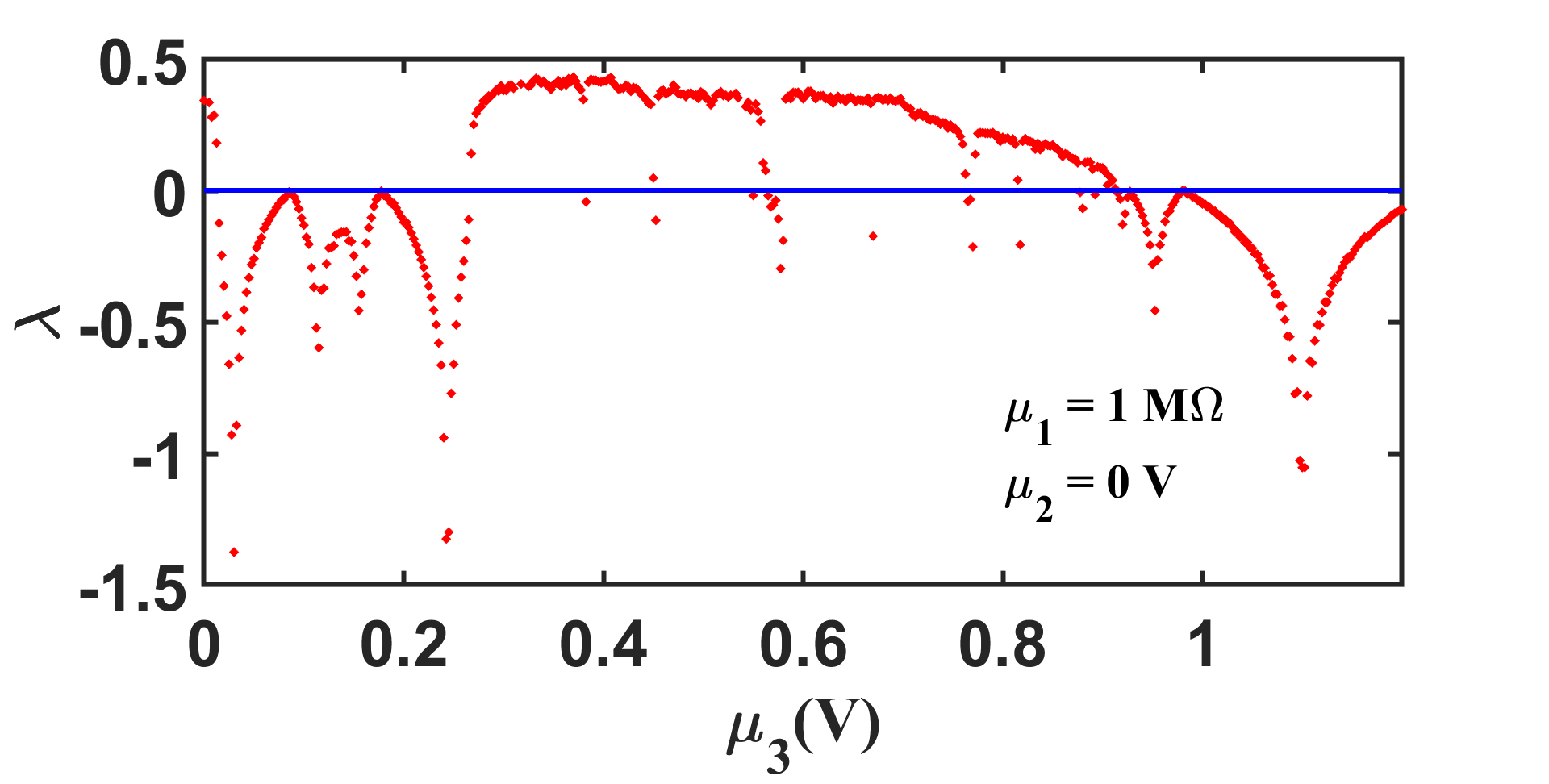}
        \caption{}
        \label{fig:LE_3}
    \end{subfigure}
    
    \caption{\small{(a) Bifurcation diagram and (b) Lyapunov exponent of the chaotic oscillator for varying $\mu_3$; $\mu_1$ = $1$ $M\Omega$,  $\mu_2$ = $0$ $V$.} }
 \label{bif_mu3}
\end{figure}

The parameters, $\mu_1$, $\mu_2$ and $\mu_3$ control the behavior of the nonlinear dynamic system which is shown in the bifurcation diagrams  Figs. \ref {fig:bif_1}, \ref {fig:bif_2} and \ref {fig:bif_3}. In Fig. \ref{fig:bif_1}, the steady-state output is plotted for varying $\mu_1$ while keeping $\mu_2$ and $\mu_3$ constant. The first 1000 iterations are ignored and the next 3000 iterations are plotted. We can clearly see the period-doubling phenomenon and chaotic regions. Similarly, Figs \ref {fig:bif_2} and \ref {fig:bif_3} show the effect of $\mu_2$ and $\mu_3$, respectively while keeping the other two parameters constant. The corresponding Lyapunov exponents are shown in Figs. \ref {fig:LE_1}, \ref {fig:LE_2} and \ref {fig:LE_3}. As can be seen from these figures, a positive value of Lyapunov exponent corresponds to a chaotic region whereas a negative value corresponds to  non-chaotic (fixed/periodic) operating region. 

\section{Logic Function generation from chaotic oscillator}
\label{sec_logic}

\begin{figure}
\centering
\includegraphics[scale=0.65]{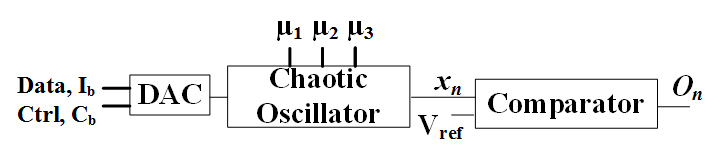}
\caption{\small{Logic gate using chaotic oscillator.}}
\label{chao_logic}
\end{figure}

The proposed chaotic oscillators can be used to build reconfigurable logic gates. A schematic for the logic gate is shown in Fig. \ref{chao_logic}. A digital-to-analog converter (DAC) converts the digital input bits to analog value which is used as the seed value ($x_0$) for chaotic oscillator. In order to enhance the design space, a control bus, $C_b$ is used along with the data bus, $I_b$. After each iteration, the analog output voltage is converted to digital voltage by using a comparator with a reference voltage, $V_{ref}$. The digital conversion is done using the following equation:
\begin{equation}
O_n=
\begin{cases}
1,&\text{if } x_{n} > V_{ref}\\
0,&\text{otherwise.}
\end{cases}
\end{equation}

In this work, $2$-input chaos-based digital gates along with $1$ control bit is used. A $3$-bit DAC is required for converting the digital input to analog voltage. The reconfigurable chaos-based logic gate can implement 16 different functions which are numbered in decimal starting from $0$ (0000) to $15$ (1111). Table \ref{tab:chaos_analogtable} demonstrates the evolution of analog voltages with different iterations. The values of the control parameters are $\mu_1$ = $0.95$ $M\Omega$, $\mu_2$ = $0$ $V$, $\mu_3$ = $0$ $V$, $C_b$ = $0$ and $V_{ref}$ = $1.25$ $V$. The functions vary with each iteration and they are represented in decimal format e.g. AND and NAND are represented by $1 (0001)$ and $14(1110)$, respectively.
\begin{table}
  \centering
  \caption{\small{Evolution of chaotic output with different iterations ($\mu_1$ = $0.95$ $M\Omega$, $\mu_2$ = $0$ V, $\mu_3$ = $0$ $V$, $C_b=0$, $V_{ref}$ = $1.25$ $V$). Functions represented in decimal value.}}
\scalebox{0.9}{
\begin{tabular}{|c|c|c|c|c|c|}
  
    \hline
    {\textbf{$x_n$} (V)} & \multicolumn{5}{c|}{\textbf{$x_{n+1}$ (V)}} \\
   
    \cline{2-6}
    & \textbf{$n=1$} & \textbf{$n=2$}& \textbf{$n=3$}&\textbf{$n=4$}&\textbf{$n=5$} \\
    \hline
    $0.1(00)$ & $0.46 (0)$ &  $1.62(1)$& $0.63(0)$ & $1.89(1)$ & $0.31(0)$ \\ \hline
    $0.757(01)$ & $1.98 (1)$ & $0.24(0)$ & $1.02(0)$ & $1.84(1)$ & $0.35(0)$\\ \hline
    $1.414(10)$ & $1.01 (0)$ & $1.86(1)$ & $0.34(0)$ & $1.33(1)$ & $1.2$(0)\\ \hline
    $2.071(11)$ & $0.19 (0)$ & $0.82(0)$ & $1.98(1)$ & $0.24(0)$ & $1.01(0)$\\ \hline
    \textbf{Func (dec) } & $4$ & $10$ & $1(AND)$ & $14(NAND)$ & $0$ \\ \hline
  \end{tabular}}
  \label{tab:chaos_analogtable}
\end{table}
As shown in \cite{hasan2020chaos}, multiple configurations  for implementing basic logic gates, enabled by nonlinear dynamics, can lead to obfuscation against power side channel attack. The proposed design can also be leveraged for the same goal. Three different configurations, distinguished by different combination of ($\mu_2,\mu_3$) are shown in  Table \ref{table:configurations} for implementing six common functions (AND, OR, XOR, NAND, NOR and XNOR). This design methodology is flexible i.e. the same logic function can be implemented in many different ways as well as reconfigurable i.e. the same circuit can be reconfigured to implement all the possible functions.

\begin{table*}

  \centering
  \caption{\small{Three different configurations for six logic functions.}}
   \scalebox{0.83}{
\begin{tabular}{|c|c|c|c|c|c|c|c|c|c|c|c|c|c|c|c|c|c|c|}
\hline
\textbf{Operation} & \multicolumn{18}{c|}{\textbf{Configurations}}\\
\cline{2-19}
&\multicolumn{6}{c|}{\textbf{Configuration 1}}&\multicolumn{6}{c|}{\textbf{Configuration 2}}&\multicolumn{6}{c|}{\textbf{Configuration 3}}\\
\cline{2-19}
&$\mu_1 (M\Omega)$&$\mu_2(V)$&$\mu_3(V)$&$C_b$&$V_{ref}(V)$&$n$&$\mu_1 (M\Omega)$&$\mu_2(V)$&$\mu_3(V)$&$C_b$&$V_{ref}(V)$&$n$&$\mu_1 (M\Omega)$&$\mu_2(V)$&$\mu_3(V)$&$C_b$&$V_{ref}(V)$&$n$\\
\hline
AND&0.97&0&0&0&1.1&3&0.94&-0.3&0&1&1.5&5&0.99&-0.3&0.3&0&1.4&7\\
\hline
OR&0.94&0&0&1&1.2&7&26&0.96&-0.3&0&0&0.7&0.99&-0.3&0.3&0&0.7&5\\
\hline
XOR&0.95&0&0&1&0.6&4&41&0.97&-0.3&0&0&0.7&0.99&-0.3&0.3&0&0.7&1\\
\hline
NAND&0.95&0&0&0&1.4&6&61&0.95&-0.3&0&1&0.5&0.99&-0.3&0.3&1&1.1&5\\
\hline
NOR&0.93&0&0&1&1.4&6&54&0.9&-0.3&0&0&1.4&0.99&-0.3&0.3&1&1.6&7\\
\hline
XNOR&0.98&0&0&1&1.6&3&63&0.92&-0.3&0&1&1.4&0.99&-0.3&0.3&0&1.6&6\\

\hline

\end{tabular}
}
\label{table:configurations}
  \end{table*}
  
\section{Enhancement of the Functionality Space}
\label{sec_func_space}
In this work we propose two new designs of chaotic oscillators (Fig. \ref{osci}) which provide a significant increase in functionality space compared to existing works. In \cite{kia2016simple}, there are three tuning parameters, bifurcation parameter ($\mu$), control bit ($C_b$) and number of iterations ($n$). Let, $N_\mu$ be the number of available levels in the chaotic region and $c$ is the number of control bits. The functionality space, $F(n)$ in the design can be written as,
\begin{equation}
F_1(n) = 2^c\times N_{\mu} \times n 
\label{space1}
\end{equation}

In \cite{shanta2018design}, threshold is varied but control bit is not used. The bifurcation parameter was changed in each iteration to extend the functionality space. If we express $N_{v_{ref}}$ as the number of comparator reference voltage levels, the functionality space can be written as,

\begin{equation}
F_2(n) = N_{v_{ref}}\times N_{\mu}^n \times n 
\label{space2}
\end{equation}

In this work, we propose two new designs. For design 1, we use chaotic oscillator $1$ from Fig. \ref {osci}a with one chaotic map. There are six tunable parameters in this design: control bits ($C_b$), bifurcation parameters ($\mu_1$, $\mu_2$ and $\mu_3$), iteration number ($n$) and reference voltage ($V_{ref}$). If we allow the bifurcation paramter to change from iteration to iteration, the resulting functionality space can be written as,

\begin{equation}
F_3(n) = N_{v_{ref}}\times2^c\times N_{\mu_1}^n  \times N_{\mu_2}^n  \times N_{\mu_3}^n \times n 
\label{space3}
\end{equation}

We can further extend the functionality space by using chaotic oscillator $2$ from Fig. \ref{osci}b. In this case, each of the forward path map and feedback path map has three independent bifurcation parameters and the functionality space can be written as,

\begin{equation}
F_4(n) = N_{v_{ref}}\times2^c\times N_{\mu_1}^{2n}  \times N_{\mu_2}^{2n}  \times N_{\mu_3}^{2n}  \times n 
\label{space4}
\end{equation}

\begin{figure}
\centering
\includegraphics[scale=0.18]{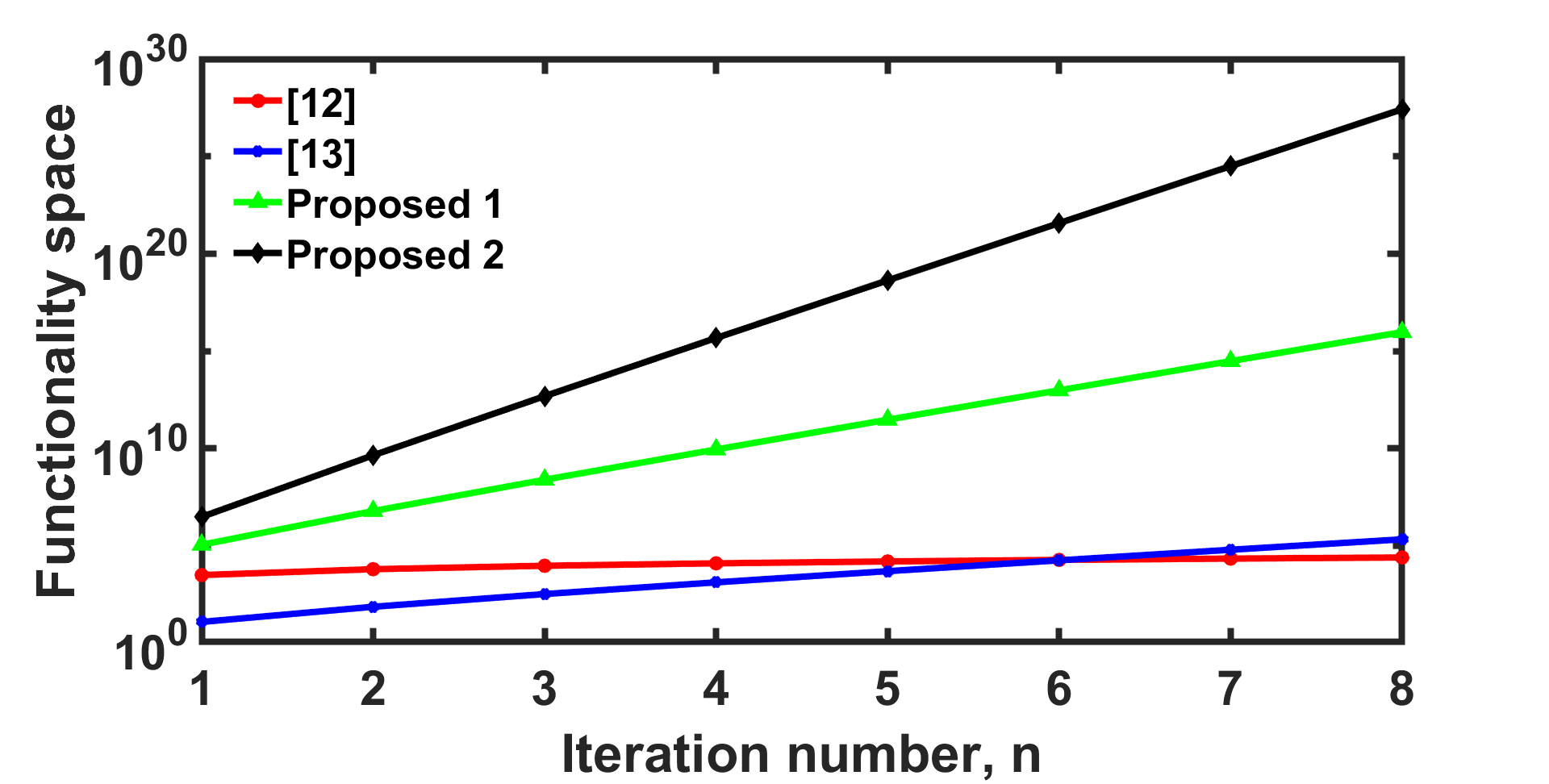}
\caption{\small{Comparison of functionality space among proposed chaotic oscillator $1$, $2$ and previous works.}}
\label{fig:func_space}
\end{figure}

The functionality spaces for equations (\ref{space1})-(\ref{space4}) are shown in Fig. \ref{fig:func_space}. This clearly shows that the functionality space of the proposed design is significantly larger compared to existing designs \cite{kia2016simple},\cite{shanta2018design} due to multiple bifurcation parameters. It should be noted that there are two more gates (bottom gates of $n$- and $p$-channel G\textsuperscript4FETs) that can also be used to further extend the design space. 

The nonlinear dynamics inherent in a chaos-based design enables developing complex (e.g. multi-input, multi-output) logic gates  using the same circuit \cite{hasan2020chaos}. The possible functions for an $n$-input logic gate is $2^{2^n}$. Hence, a 3-input gate will have 256 possible functions. For obfuscation applications, it is highly desirable to have a large functionality space so that multiple reliable configurations can be chosen for implementing a single function. This can be leveraged to ensure the security of chaos-based computing systems against side-channel power analysis based attacks \cite{hasan2020chaos}, \cite {8383903}. Previously, this was achieved using many iterations but this poses a problem since the resulting gate becomes increasingly unreliable. As shown in Fig. \ref{fig:func_space}, the proposed design can achieve a very large functionality space using very few iterations. 

Though a single chaos-gate has a larger overhead compared to conventional CMOS gates, this kind of multi-input arbitrary logic function can dramatically reduce the overhead of the overall design. For example, let's consider a three input one output function, $Y=ABC+ \overline{A}B\overline{C}+A\overline{B}\, \overline{C}  + \overline{A}\,\overline{B}C.$ If we only use two-input logic gates, we have to use 14 gates whereas a single three-input one-output chaos gate can implement the same function. This approach can significantly reduce the overhead for implementing complex logic functions. Moreover, as shown in \cite {8383903}, oftentimes we only have to replace a small percentage of total gates with chaos-based implementation to benefit from the resulting security advantages.

\section{Conclusion}
In this paper, a novel nonlinear chaotic map circuit is introduced using a negative differential resistance circuit made of two complementary SOI four-gate transistors. Two novel chaotic oscillator designs are proposed and 
the resulting periodic and chaotic regions are demonstrated using bifurcation diagrams and Lyapunov exponent. A methodology for designing flexible and reconfigurable logic gate is outlined and three representative configurations for implementing basic logic gates are shown. Compared to earlier works, significant enhancements in functionality space have been demonstrated. The proposed designs can be utilized to reduce hardware overhead by reliably implementing complex logic functions. The enhanced space can be utilized for mitigating multiple hardware security problems. 
\label{sec_con}

\bibliographystyle{IEEEtran}

\bibliography{DCAS_chaos.bib}




\end{document}